\documentclass[aps,pra,a4paper,twocolumn,showpacs,floatfix]{revtex4}
\usepackage[dvips]{epsfig}
\begin{document}
\title{Ultracold quantum dynamics: spin-polarized K + K$_{2}$
collisions\\ with three identical bosons or fermions}
\author{Goulven Qu\'{e}m\'{e}ner, Pascal Honvault, Jean-Michel
Launay,}

\affiliation{UMR 6627 du CNRS, Laboratoire de Physique des Atomes,
Lasers, Mol\'ecules et Surfaces, Universit\'e de Rennes, France}

\author{Pavel Sold\'{a}n, Daniel E. Potter, and Jeremy M. Hutson}

\affiliation{Department of Chemistry, University of Durham, South
Road, Durham, DH1~3LE, England}
\date{\today}

\begin{abstract}
We have developed a new potential energy surface for
spin-polarized K($^2$S) + K$_{2}(^3\Sigma^+_u)$ collisions and carried out quantum
dynamical calculations of vibrational quenching at low and
ultralow collision energies for both bosons $^{39}$K and $^{41}$K
and fermions $^{40}$K. At collision energies above about 0.1 mK
the quenching rates are well described by a classical Langevin
model, but at lower energies a fully quantal treatment is
essential. We find that for the low initial vibrational state
considered here ($v=1$), the ultracold quenching rates are {\it
not} substantially suppressed for fermionic atoms. For both bosons
and fermions, vibrational quenching is much faster than elastic
scattering in the ultralow-temperature regime. This contrasts with
the situation found experimentally for molecules formed via
Feshbach resonances in very high vibrational states.
\end{abstract}

\maketitle

\font\smallfont=cmr7

\section{Introduction}

Dilute gases of alkali-metal atoms are a rich source of novel
physical phenomena. Bose-Einstein condensates (BECs) were first
created in such gases in 1995 \cite{And95,Ket95,Hul95} and have
been the subject of intense exploration ever since. Further
possibilities were opened up by the achievement of Fermi
degeneracy in 1999 \cite{Jin99}. Among the alkali-metal atoms,
lithium \cite{Hul95,Hul01} and potassium \cite{Jin99,Ing01} have a
special status because both bosonic and fermionic isotopes are
available and both Bose-Einstein condensation and Fermi degeneracy
have been achieved. The present paper will focus on potassium.


In recent years, much interest has focussed on the interactions
between atoms and the formation of molecules in ultracold gases.
Donley {\it et al.}\ \cite{Wiem02} showed that it is possible to
form dimers of bosonic atoms such as $^{85}$Rb by magnetic tuning
from an atomic to a molecular state in the vicinity of a Feshbach
resonance. However, molecules formed in this way in an atomic BEC
proved to be short-lived (with lifetimes of milliseconds) because
of atom-molecule and molecule-molecule collisions
\cite{Wiem02,Grimm03c,Remp04,Kett04}: the molecules are formed in
the highest vibrational state that exists in the two-body
potential well, and any collision that changes the vibrational
state releases enough energy to eject both collision partners from
the trap.

In summer 2003, Regal {\it et al.}\ \cite{Jin03b} succeeded in
forming ultracold diatomic molecules in a Fermi-degenerate gas of
$^{40}$K atoms by ramping the magnetic field through a Feshbach
resonance. Such molecules are composite bosons. However, these too
turned out to be short-lived (lifetime $< 1$ ms). Finally, at the
end of 2003, a long-lived molecular BEC was created using the same
technique \cite{Jin03a} with a different Feshbach resonance. Such
condensates have also been formed from dimers of $^{6}$Li
\cite{Grimm03a,Kett03a,Salomon04}.

The use of fermionic isotopes appears to be crucial for the
production of long-lived molecular condensates. The inelastic
collisions that cause trap loss for molecules formed from bosonic
atoms are sometimes suppressed for molecules formed from fermionic
atoms \cite{Jin04a,Sal03,Hulet03a,Grimm03b}. At magnetic fields
where the atom-atom scattering length is large and positive ($a >
1000$ a$_{0}$), molecular lifetimes longer than 100 ms can be
achieved. Petrov {\it et al.}\ \cite{Gora03} have explained the
difference in collisional properties of dimers formed from bosonic
and fermionic atoms in terms of the symmetries of the allowed
wavefunctions. However, their derivation applies only to dimers in
Feshbach resonance states and not to deeply bound molecular
states.

In a previous study \cite{Cvit04}, we investigated
ultralow energy collisions between spin-polarized Li atoms and
Li$_2$ dimers, with the dimers in low-lying vibrational bound
states. Our results showed no systematic differences in
vibrational quenching between the bosonic $^{7}$Li and fermionic
$^{6}$Li cases. This supports the conclusion that the
suppression of inelastic collisions in the fermionic case requires
{\it both} fermion symmetry \cite{Gora03} {\it and} the long-range
nature of the molecules in Feshbach resonance states \cite{Bur03}.

In the present paper, we study ultralow energy K + K$_{2}$
collisions involving three equivalent nuclei. We construct a new
\textit{ab initio} potential energy surface for the lowest
spin-polarized electronic state of the potassium trimer
($1^{4}A_{2}'$) and investigate both bosonic ($^{39}$K, $^{41}$K)
and fermionic ($^{40}$K) cases. We perform quantum-mechanical
scattering calculations at energies down to 1~nK. Elastic,
inelastic and rearrangement processes are considered. Our quantum
dynamical results show that, for all three systems, vibrational
relaxation is more efficient than elastic scattering, as in our
previous studies with $^{23}$Na \cite{Sol02,Quem04} and $^{6,7}$Li
\cite{Cvit04}. As in the case of lithium, there are no systematic
differences in quenching rates for potassium dimers formed from
bosonic and fermionic atoms.

\section{Potential energy surface of K$_3(1^{4}A_2')$}

We have carried out \textit{ab initio} calculations on K$_3$ using
a single-reference restricted open-shell variant \cite{KHW93} of
the coupled-cluster method \cite{Cizek} with single, double and
non-iterative triple excitations [RCCSD(T)]. We used the
small-core ECP10MWB effective core potential (ECP) of Leininger
{\it et al.}\ \cite{ECP} together with the medium-sized valence
basis sets of Sold\'{a}n \textit{et al.}\ \cite{Soldan03}. The
quasirelativistic \cite{WB} ECP treats the 1s$^2$ electrons as
core and the 3s$^2$ 3p$^6$ 4s electrons as valence. The
valence basis set for K was used in uncontracted form. The
resulting atomic polarizability (294.2 $a_0^3$) is in excellent
agreement with the experimental value
\cite{Polar1} ($292.8 \pm 6.1$ $a_0^3$). \\

For interpolation purposes, the three-atom interaction potential
was decomposed into a sum of additive and non-additive
contributions,
\begin{equation}
\label{Eq1} V_{\rm trimer}(r_{12},r_{23},r_{13}) = \sum_{i<j}
V_{\rm dimer}(r_{ij}) + V_{3}(r_{12},r_{23},r_{13}).
\end{equation}
The full counterpoise correction of Boys and Bernardi \cite{BSSE}
was employed to compensate for basis set superposition error in
both dimer and trimer calculations. All the \textit{ab initio}
calculations were performed using the MOLPRO package
\cite{MOLPRO}.

\begin{figure} [htbp]
\begin{center}
\rotatebox{270}{ \resizebox{8cm}{!} {\includegraphics{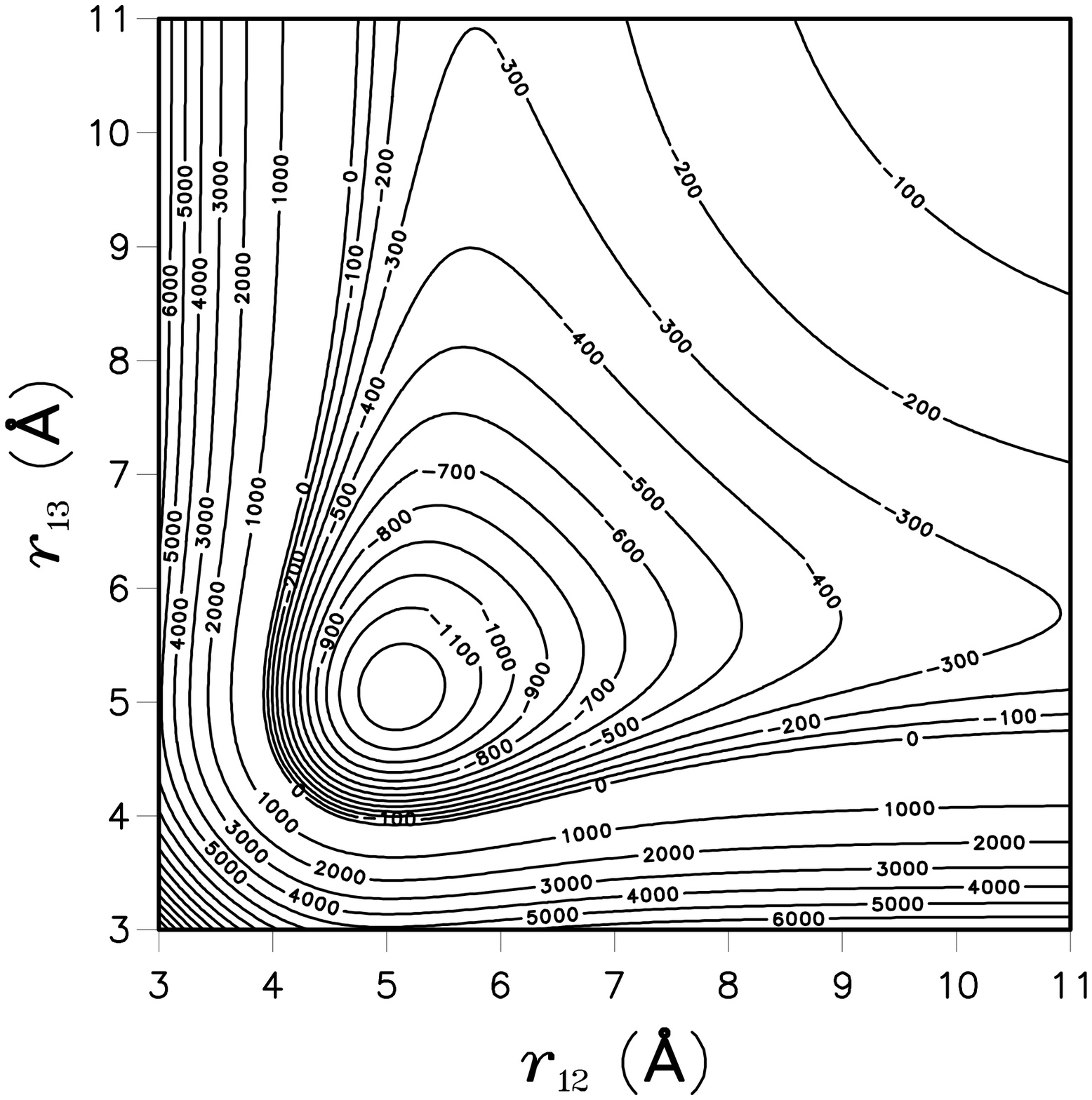}}}
\rotatebox{270}{\resizebox{8cm}{!} {\includegraphics{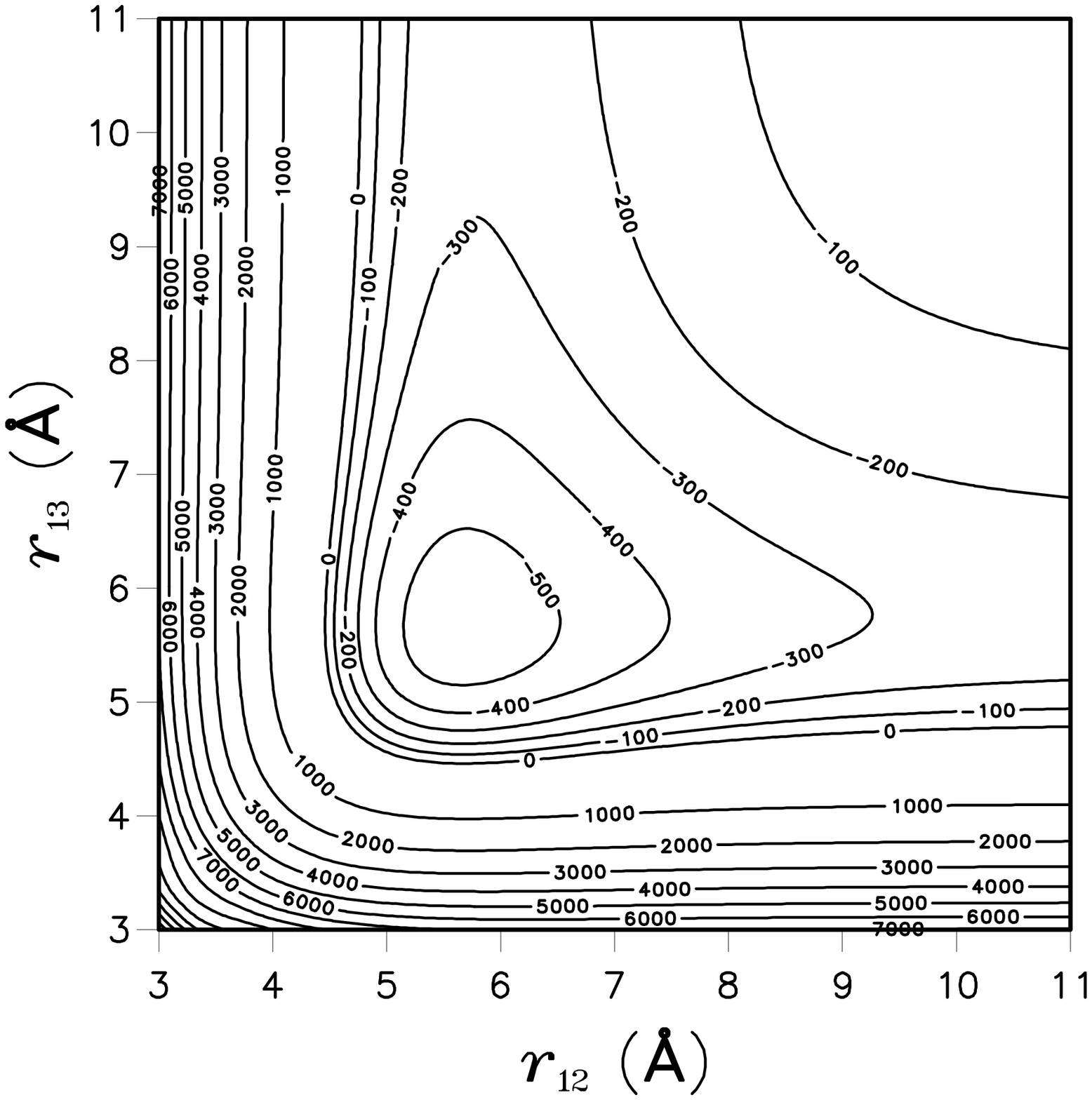}}}
\caption{Cuts through the K$_3$ quartet surface in valence
coordinates. Upper panel: cut for a bond angle of 60$^\circ$,
showing the global minimum at $-1269$ cm$^{-1}$ and 5.09 \AA.
Lower panel: cut at collinear geometries; the collinear minimum is
at $-565$ cm$^{-1}$ and 5.68 \AA. Contours are labelled in
cm$^{-1}$.} \label{k3pot}
\end{center}
\end{figure}

The dimer interaction energy $V_{\rm dimer}(r)$ was first
calculated on an irregular grid of 42 points at interatomic
distances between 2.1 \AA\ and 14.0 \AA. The potential energy
curve was generated using the modified 1D reciprocal-power
reproducing kernel Hilbert space (RP-RKHS) interpolation method
\cite{Ho96,Ho00}. The interpolation was done with respect to
$r^{2}$ using RP-RKHS parameters $m=2$ and $n=3$. Beyond 14 \AA\
the potential energy curve was thus extrapolated to the form
\begin{equation}
\label{Eq2} V_{\rm dimer}(r) = -\frac{C_{6}}{r^{6}} -
\frac{C_{8}}{r^{8}} - \frac{C_{10}}{r^{10}}.
\end{equation}
The long-range coefficients $C_{6}$ and $C_{8}$ were kept fixed to
the published values of $3.897\times10^{3}$ $E_{h} a_0^{6}$ and
$4.2 \times 10^{5}$ $E_{h} a_0^{8}$ respectively
\cite{Der99,Der03}. The value of the ``free'' long-range
coefficient $C_{10}$ was then determined from the corresponding
RP-RKHS coefficients \cite{Sol00}, and was found to be $2.0243
\times 10^{10}$ $E_{h}\,a_0^{10}$, which compares very well with
$2.0294 \times 10^{10}$ $E_{h}\,a_0^{10}$ from Ref.\
\onlinecite{Der03}. Values of $D_{\rm e}=252.6$ cm$^{-1}$ and
$r_{\rm e} = 5.79$ \AA\ calculated from the resulting curve are
also in good agreement with experimental results on $^{39}$K$_{2}$
($D_{\rm e} = 252.74 \pm 0.12$ cm$^{-1}$, $r_{\rm e} = 5.7725(20)$
\AA\ \cite{K1,K2}).

For quantum dynamics calculations, it is very important to have a
potential energy function that can be interpolated smoothly (and
without oscillations) between {\it ab initio} points. Oscillations
often arise in low-energy regions if one or more points have much
higher energies than those surrounding them. Some experimentation
was needed to find a coordinate system in which interpolation
could be carried out without problems. Jacobi and bond angle /
bond length coordinates were rejected because they do not lend
themselves to symmetrization, and hyperspherical coordinates
proved unsuitable because some combinations of grid points produce
geometries with atoms very close together and correspondingly high
energies. In the end, we chose to calculate the potential grid in
pure bond length coordinates $(r_{12},r_{23},r_{31})$. This has
the advantage that points that are related by symmetry have
coordinates that are also simply related.

The trimer interaction energy $V_{\rm trimer}$ was calculated at
325 points on a 3D grid covering the range of interatomic
distances from 3.5 \AA\ to 10.5 \AA\ with step 0.5 \AA. To avoid
duplication, only points with $r_{12} \le r_{13} \le r_{23}$ are
required. To meet geometrical constraints, all grid points must
satisfy the triangular inequality $|r_{12}-r_{13}|\leq r_{23} \leq
r_{12}+r_{13}$.  The distance $r_{23}$ was permitted to extend
beyond 10.5 \AA. The final grid consisted of 205 C$_{\rm 2v}$
points, including 15 D$_{\rm 3h}$ points and 120 C$_{\infty{\rm
v}}$ points; the latter include 15 D$_{\infty{\rm h}}$ points.
Each calculation was carried out using the full symmetry allowed
by MOLPRO. The non-additive energies $V_3$ were extracted from the
trimer interaction energies using Eq.\ (\ref{Eq1}).

For low-energy scattering calculations it is important to have an
interaction potential that behaves correctly at long range. The
RP-RKHS interpolation procedure in 1 dimension allows this as
decribed above. However, multidimensional RP-RKHS interpolation
always gives a potential that extrapolates beyond the points as a
simple product of inverse powers in the different coordinates.
The leading long-range terms in the non-additive energy
are the third-order dipole-dipole-dipole (DDD) \cite{ATM} and
dipole-dipole-quadrupole (DDQ) \cite{Bell} terms given by
\begin{equation}
\label{DDD3} V^{\rm DDD}_3=3Z^{(3)}_{111}
\frac{1+3\cos{\phi_{3}}\cos{\phi_{1}}\cos{\phi_{2}}}
{r_{12}^3r_{23}^3r_{13}^3}
\end{equation}
and
\begin{equation}
\label{DDQ3a} V^{\rm DDQ}_{3} = Z^{(3)}_{112}\,(W^{123} + W^{231}
+ W^{312})
\end{equation}
where
\begin{widetext}
\begin{eqnarray}
W^{ijk}&=& \frac{3}{16r_{jk}^4r_{ik}^4r_{ij}^3}
\left[9\cos{\phi_{k}}-25\cos 3\phi_k +6\cos(\phi_i-\phi_j) \times
(3+5\cos 2\phi_k)\right] \label{DDQ3b}
\end{eqnarray}
and $\phi_i$ is the bond angle at atom $i$. It may be noted that
the DDD term vanishes on a seam in the angular space and the DDQ
term vanishes at all linear geometry configurations. Damped
versions of these terms were therefore subtracted from the total
non-additive energy $V_3$ before interpolation to give a quantity
$V'_3$,
\begin{equation} V_3' = V_3 - f_{\rm damp} [V^{\rm DDD}_3
+ V^{\rm DDQ}_3].\end{equation} The coefficients $Z^{(3)}_{111}$
and $Z^{(3)}_{112}$ were taken to be $2.72\times10^{5}$
$E_{h}\,a_0^{9}$ and $5.11\times10^{6}$ $E_{h}\,a_0^{11}$
respectively \cite{PT}. The damping function serves to prevent the
non-additive energy exploding at short range, and was chosen to
have a product form, $f_{\rm damp}(r_{12},r_{23},r_{13}) =
f(r_{12})\,f(r_{23})\,f(r_{13})$, where
\begin{eqnarray}
f(r)  =& \exp{\left[-\left(k_3/r - 1\right)^2\right]} &\ \ 0<r<k_3
\\ =& 1 &\ \ r \geq k_3
\end{eqnarray}
with the cut-off parameter $k_3=8.0$ \AA.

The leading term of the multipole asymptotic expansion of $V_{3}'$
is the fourth-order dipole-dipole-dipole term (DDDD), which has a
more complicated (unfactorizable) form \cite{Bade},
\begin{eqnarray}
V^{\rm DDDD}_{3} &=& -\frac{45}{64} Z^{(3)}_{1111}
\left[\frac{1+\cos^2\phi_{1}}{r_{12}^6r_{13}^6}+
\frac{1+\cos^2\phi_{2}}{r_{12}^6r_{23}^6}+
\frac{1+\cos^2\phi_{3}}{r_{13}^6r_{23}^6}\right]. \label{DDD4}
\end{eqnarray}
The coefficient $Z^{(3)}_{1111}$ is not yet known, so this term
cannot be subtracted out. However, the term is negative at all
geometries, so it can be eliminated by defining $V_{3}'' = g
\times V_{3}'$, where
\begin{equation}
g =
\frac{r_{12}^3r_{23}^3r_{13}^3}{(1+\cos^2\phi_{1})\,r_{23}^{6}+
(1+\cos^2\phi_{2})\,r_{13}^{6}+(1+\cos^2\phi_{3})\,r_{12}^6}.
\label{V3pp}
\end{equation}
\end{widetext}
The leading asymptotic term of the function $V_{3}''$ now has the
form $-\hbox{constant}\times r_{12}^{-3}r_{23}^{-3}r_{13}^{-3}$
and is suitable for an ``isotropic'' extrapolation of the type
that results from a multidimensional RP-RKHS interpolation. The
function $V_{3}''$ was interpolated using the fully symmetrized 3D
RP-RKHS interpolation method \cite{Hig00}. The interpolation was
done with respect to the reduced coordinate $(r/S)^3$ and with
parameters $S=10.0$ \AA, $m=0$, $n=2$ in each interatomic
distance. The original potential is then rebuilt as
\begin{equation}
V_3 = \frac{1}{g}\,V_3'' + f_{\rm damp} \left[V^{\rm DDD}_3 +
V^{\rm DDQ}_3\right]. \label{Vorig}
\end{equation}

The final potential for quartet K$_3$, $V_{\rm trimer}$, has a
global minimum at $-1269$ cm$^{-1}$ at an equilateral (D$_{\rm
3h}$) geometry $r_{12}=r_{13}=r_{23}=5.09$ \AA. There is a shallow
secondary minimum at $-565$ cm$^{-1}$ at a linear D$_{\infty{\rm
h}}$ geometry with $r_{12}=r_{13}=5.68$ \AA. Two cuts through the surface are
shown as contour plots in Fig.\ \ref{k3pot} for values of the
valence angle 60$^{\circ}$ and 180$^{\circ}$.

\section{Quantum Scattering theory}
\subsection {Method}

We have performed three-dimensional quantum dynamical calculations
for K + K$_2$ including reactive scattering for total angular
momenta $J=0 - 5$. A time-independent formalism (which is the most
appropriate choice for ultralow energy scattering) was used. The
configuration space is divided into an inner and an outer region
depending on the atom--diatom distance. In the inner region,
typically for hyperradius smaller than $\rho_{\rm max}$ = 60
a$_0$, we use a formalism based on body-frame democratic
hyperspherical coordinates \cite{launay89,hon04} which has
previously proved successful in describing atom-diatom insertion
reactions such as N($^2$D) + H$_2$ $\rightarrow$ NH + H
\cite{nh2jcp99,balu02} and O($^1$D) + H$_2$ $\rightarrow$ OH + H
\cite{oh2jcp01,oh2prl01}. These coordinates were also used in our
recent work on Na + Na$_2$ \cite{Sol02,Quem04} and Li + Li$_2$
\cite{Cvit04}.

At each hyperradius $\rho$, we determine a set of eigenfunctions
of a fixed-hyperradius reference hamiltonian $H_0=T+V$ by
expanding the wavefunction in a set of pseudohyperspherical
harmonics. The reference hamiltonian incorporates the kinetic
energy $T$ arising from deformation and rotation around the axis
of least inertia and the potential energy $V$. A typical set of
eigenvalues for K$_3$ is shown in Fig.\ \ref{SPAGK3}. At small
hyperradius, the adiabatic states in each sector span a large
fraction of configuration space and allow for atom exchange. The
scattering wave function is expanded on this set of hyperspherical
adiabatic states. This yields a set of close-coupling equations,
which are solved using the Johnson--Manolopoulos log-derivative
propagator \cite{mano86}.

In the outer region, we use the standard Arthurs-Dalgarno
formalism \cite{arthurs60} based on Jacobi coordinates and assume
that the atom--molecule interaction can be described by an
isotropic potential $U(R)$. We thus assume that no inelastic
scattering occurs in the outer region. The radial channel
wavefunctions in this region are solutions of the equation (in
atomic units)
\begin{equation}
\Bigl( - \frac{1}{2\mu} \frac{d^2}{dR^2} + \frac{l(l+1)}{2\mu R^2}
+ U(R) \Bigr) F(R) = \frac{k^2}{2\mu} F(R), \label{radial}
\end{equation}
where $\mu$ is the K--K$_2$ reduced mass and $k$ and $l$ are
respectively the channel wavenumber and the orbital angular
momentum quantum number. The potential $U(R)$ behaves
asymptotically as $-C_{\rm K-K_2}/R^6$ with a coefficient $C_{\rm
K-K_2}$ = 9050 $E_ha_0^6$. 
Equation~(\ref{radial}) is solved by a finite difference method in
the range 60--10000 a$_0$. The matching of the inner and outer
wavefunctions is performed on a boundary which is an hypersphere
of radius $\rho=60\ a_0$. This yields the reactance $K$,
scattering $S$ and transition $T$ matrices.

\subsection {Cross sections and Wigner laws}

At ultralow energies, for {\it both} bosonic and fermionic
systems, only $T_{ii}$ elements with a relative atom--diatom
orbital angular momentum $l=0$ contribute. We obtain elastic and
quenching cross sections from the diagonal elements $T_{ii}$ of
the transition $T$ matrix,
\begin{equation}
\sigma_E = \frac{\pi}{k^2} |T_{ii}|^2; \hskip 5mm \sigma_Q =
\frac{\pi}{k^2} (1-|1-T_{ii}|^2). \label{sigma}
\end{equation}
The complex scattering length can be written as
\begin{equation}
a = \frac{1}{2i} \lim_{k \rightarrow 0} \Bigl( \frac{T_{ii}}{k}
\Bigr). \label{limit}
\end{equation}
The elastic and quenching cross sections at threshold can
therefore be written respectively in terms of the square modulus
of the scattering length and the imaginary part of the scattering
length,
\begin{equation}
\sigma_E = 4\pi|a|^2; \hskip 5mm \sigma_Q = - \frac{4\pi}{k}
Im(a).
\end{equation}
The elastic and quenching rate coefficients $K_E$ and $K_Q$ are
obtained by multiplying the cross sections by the atom--diatom
relative velocity, $k/\mu$ in atomic units,
\begin{equation}
K_E = \frac{4\pi |a|^2}{\mu} k; \hskip 5mm K_Q = -
\frac{4\pi}{\mu} Im(a).
\end{equation}
At ultralow collision energy, where only $l=0$ is involved, the
elastic rate coefficient vanishes as the collision energy
decreases and the quenching rate coefficient becomes constant.

The Wigner threshold laws \cite{wigner48} for an orbital angular
momentum $l$ give the following dependences on the collision
energy for partial elastic and quenching cross sections,
\begin{equation}
\sigma_E^l  \sim E_{\rm coll}^{2l}; \hskip 5mm \sigma_Q^l \sim
E_{\rm coll}^{l-1/2}. \label{wcross}
\end{equation}
The energy-dependence for partial elastic and quenching rate
coefficients is obtained by multiplying these cross sections by
$(2E_{\rm coll}/ \mu)^{1/2}$,
\begin{equation}
K_E^l  \sim E_{\rm coll}^{2l+1/2}; \hskip 5mm K_Q^l \sim E_{\rm
coll}^{l}.
\end{equation}

We also need state-to-state cross sections for the calculation of
rotational distributions. For partial wave $J$, they are generated
from the $T$ matrix using the standard formula,
\begin{widetext}
\begin{equation} \sigma_{vj \rightarrow
v'j'}=\frac{\pi}{(2j+1)k_{vj}^2} \sum_{J=0}^{\infty}(2J+1)
\sum_{l=|J-j|}^{J+j} \sum_{l'=|J-j'|}^{J+j'} |T^{J}_{v'j'l',
vjl}|^2. \label{limit2}
\end{equation}

In this paper, we present for the first time differential cross
sections (DCS) for an atom-diatom system at ultralow collision
energy. The state-to-state magnetically averaged DCS is given by
\begin{equation}
\frac{d\sigma_{vj \rightarrow v'j'}}{d\omega} =
\frac{1}{4k^2_{vj}(2j+1)} \sum_{m,m'}\left|\sum_{J}(2J+1)
d^J_{m',m}(\theta_{\rm cm}) T^{J}_{v'j'm', vjm} \right|^2
\label{dcs1}
\end{equation}
\end{widetext}
where $d^J_{m',m}$ is a Wigner reduced rotation matrix element.

In Eq.\ (\ref{dcs1}), $\theta_{\rm cm}$ is the scattering angle in
the center-of-mass (CM) coordinate system. $\theta_{\rm cm}=0
^\circ $ is defined as the direction of the CM velocity vector of
initial K atoms and corresponds to forward scattering for the K
products (and backward scattering for the K$_2$ products).
Backward scattering of the K products (forward scattering for the
K$_2$ products) thus corresponds to $\theta_{\rm cm}=180 ^\circ$.

\subsection{Symmetries}

We are interested in elastic scattering and vibrational
relaxation, which have different roles in the formation and
stability of a molecular BEC. Elastic collisions are favorable for
evaporative cooling towards condensation, whereas inelastic
collisions provide a trap loss mechanism. For evaporative cooling,
knowledge of the ratio of elastic to non-elastic rate coefficients
is crucial.

Here the system is composed of three indistinguishable atoms
$^{39}$K, $^{40}$K or $^{41}$K and the K$_3$ potential energy
surface is barrierless. We thus have to take into account two
different collisional processes: elastic collision K + K$_2(v,j)$, 
and vibrational relaxation K + K$_2(v,j) \rightarrow$ K + K$_2(v',j')$. 
For this latter process, the rovibrational energy $E_{v',j'}$ of the product
molecule is smaller than that of the initial molecule.

We consider three atoms K in their stretched spin states, with
$F=F_{\rm max}=S+I$ and $M_F=F$. For $S=1/2$, this requires $M_S =
+1/2$ and so the electronic spin wavefunction is symmetric with
respect to exchange of nuclei. $^{39}$K and $^{41}$K atoms have
$I=3/2$, so $M_I=+3/2$ in the spin-stretched state and the nuclear
spin wavefunction is also symmetric. Since the total spin $F$ is
an integer, these atoms are composite bosons. For $^{40}$K, $I=4$
so that $F$ is half-integer and the atoms are composite fermions.
For the spin-stretched state of $^{40}$K, $F=F_{\rm max}=I+S=9/2$
and $M_F=M_{\rm max} = +9/2$. In this work, all magnetic
interactions are neglected, so that the electronic and nuclear
spin orientations are unchanged during the collision.

The spatial part of the {\it electronic} wavefunction for the
$1^4A_2^\prime$ state is antisymmetric with respect to exchange of
nuclei. As a consequence, for $^{39}$K or $^{41}$K the spatial
part of the nuclear wavefunction has to be symmetric, whereas for
$^{40}$K it has to be antisymmetric. In our quantum scattering
code, this is achieved by selecting pseudohyperspherical harmonics
in the basis sets for bosons and fermions to give the correct
symmetry for the nuclear wavefunction. The adiabatic states in
each sector are obtained by a variational expansion on a basis of
hyperspherical harmonics with $A_1$ symmetry (of the complete
nuclear permutation group $S_3$) for bosons and $A_2$ symmetry for
fermions. They are respectively fully symmetric and fully
antisymmetric with respect to particle permutations to account for
the indistinguishability of atoms.

\subsection{Partial waves}

The behavior at the lowest collision energies is more subtle in
the molecular case than in the atomic case. In atom-atom
collisions, the fundamental difference between bosons and fermions
is seen in the behaviour of cross sections. For two identical
spin-polarized bosons, the elastic cross section becomes constant
as the collision energy goes to zero because only the s-wave
($l=0$) contributes. In contrast, for two fermions, the elastic
cross section decreases as the collision energy decreases because
there is no s-wave and only the p-wave ($l=1$) contributes at
ultralow energy.

In atom-molecule collisions, by contrast, $l=0$ is allowed for
{\it both} bosons and fermions. For three identical atoms, only
one partial wave is involved at ultralow energy: $J=0^+$ for
bosons and $J=1^-$ for fermions. These both include $l=0$.

The channels contributing to each partial wave $J^\Pi$ up to $J=3$
are shown for bosons in Table~\ref{tab1}. The total mechanical
angular momentum {\bf J = j + l} is the vector sum of the diatomic
rotational angular momentum {\bf j} and the orbital angular
momentum {\bf l}. In this study, we consider initial $^{39}$K$_2$
or $^{41}$K$_2$ molecules initially in rovibrational state
$(v=1,j=0)$, so the parity $\Pi$ is $(-1)^{j+l}= (-1)^l= (-1)^J$.
For $^{40}$K$_2$, only odd $j$ is allowed so we consider the
initial state $(v=1,j=1)$. The partial waves for fermions (with
contributing channels shown in Table~\ref{tab2}) can be separated
into two categories depending on the lowest value of $l$ (or
$\Omega$, which is the projection of the total angular momentum
$J$ onto the axis of least inertia in the corresponding
hyperspherical basis set). The parity-favored partial waves
include $\Omega=0$ and the parity is the same as for the boson
case. In contrast, for the parity-unfavored partial waves the
$\Omega=0$ component is forbidden (as is the lowest value of $l$)
and the lowest value of $\Omega$ is 1. The parity-favored and
unfavored partial waves have parity $(-1)^J$ and $(-1)^{J+1}$
respectively.

\begin{table}
\caption{Channels contributing to partial waves $J^{\Pi}$, showing
allowed values of the orbital angular momentum $l$ 
for a rovibrational state $(v,j=0)$ of $^{39}$K$_2$ or
$^{41}$K$_2$ molecules formed from bosonic atoms.} \vskip 0.3cm
\begin{center}
\label{tab1}
\begin{tabular}{|c|c|c|c|}
\hline
$0^+$ & $1^-$ & $2^+$ & $3^-$ \\
$l=0$  & $l=1$ & $l=2$ & $l=3$ \\
\hline
\end{tabular}
\end{center}
\end{table}

\begin{table}
\caption{Channels contributing to partial waves $J^{\Pi}$, showing
allowed values of the orbital angular momentum $l$ 
for a rovibrational state $(v,j=1)$ of a $^{40}$K$_2$ molecule
formed from fermionic atoms.} \vskip 0.3cm
\begin{center}
\label{tab2}
\begin{tabular}{|c|c|c|c|}
\hline
$0^+$ & $1^-$ & $2^+$ & $3^-$ \\
$l=1$  & $l=0,2$ & $l=1,3$ & $l=2,4$ \\
\hline
         & $1^+$ & $2^-$ & $3^+$ \\
         & $l=1$ & $l=2$ & $l=3$ \\
\hline
\end{tabular}
\end{center}
\end{table}


\subsection{Convergence and computer requirements}

K + K$_2$ collisions are the most difficult and demanding ever
studied in quantum dynamics calculations, because all three atoms
are heavy and the potential well is deep. Careful attention has
been devoted to the convergence of the calculations. There are
three crucial parameters for convergence: the sizes of the basis
sets, the size of each sector and the asymptotic matching
distance.

First, the number of fixed-$\rho$ eigenstates included in solving
the coupled equations in each sector must be large enough for
convergence. It is essential to include many closed channels. For
both bosonic and fermionic systems, this number, which is also the
number of coupled equations, increases about from 250 for $J=0$ to
1411 for $J=5$. The time taken to solve the coupled equations
varies respectively from 2 minutes to 4 hours per collision energy
on a Power4 P690 IBM computer.
The fixed-$\rho$ eigenstates are expanded in a
pseudo-hyperspherical harmonic basis built from trigonometric
functions, truncated at $\Lambda_{\rm max}$, the maximum value of
the grand angular momentum. $\Lambda_{\rm max}$ varies from 198
(867 harmonics) at small hyperradius to 558 (6625 harmonics) at
large hyperradius. The calculation to build the basis sets takes
180 hours and produces an output binary file of 40 Gigabytes which
contains all information for the close-coupling code.

Secondly, the size of each sector must be small enough to give
converged results. For both bosons and fermions, two different
sector sizes were used: 0.025 a$_0$ from $\rho=8.0$ to  26.0 a$_0$
and 0.05 a$_0$ from 26 a$_0$ to the matching distance 60 a$_0$.
This yields 1400 sectors, which is about 4 times larger than were
used for the case of Na$_3$ \cite{Sol02,Quem04}.

Thirdly, the matching distance $\rho_{\rm max}$, where the
adiabatic states are projected onto the arrangement channels, must
be chosen to describe correctly K$_2$ molecules in the vibrational
states required (in even $j$ states for bosons and odd $j$ states
for fermions). We have taken $\rho_{\rm max}=60$ a$_0$, which is
larger than the value of 50 a$_0$ for the Na$_3$ case. For bosons,
the hyperspherical wavefunction was projected onto a set of
$^{39}$K$_2$ or $^{41}$K$_2$ rovibrational functions with $j_{\rm
max} = (76,70,66,60,56,50,42,34,24,2)$ for $v=0,\ldots,9$. For
fermions $^{40}$K$_2$, rovibrational functions with $j_{\rm max} =
(77,71,67,61,55,49,43,35,25,7)$ were used.

At collision energies below 1 $\mu$K, only the partial wave that
includes $l=0$ contributes to cross sections ($J=0^+$ for bosons
and $J=1^-$ for fermions). However at higher energies, other
partial waves contribute. In this work, for the bosonic cases we
have converged the elastic and quenching cross sections for
collision energies up to 10 mK by including partial waves up to
$J=5$. Calculations for the fermionic system $^{40}$K +
$^{40}$K$_2$ are converged up to 100 $\mu$K by including partial
waves up to $J=2$. We have considered only three values of $J$ for
fermions because the calculations are twice as expensive for
fermions as for bosons (because both parity-favored and
parity-unfavored partial waves contribute in the fermion case).

\begin{figure}[h]
\begin{center}
\includegraphics*[height=6cm,width=8cm]{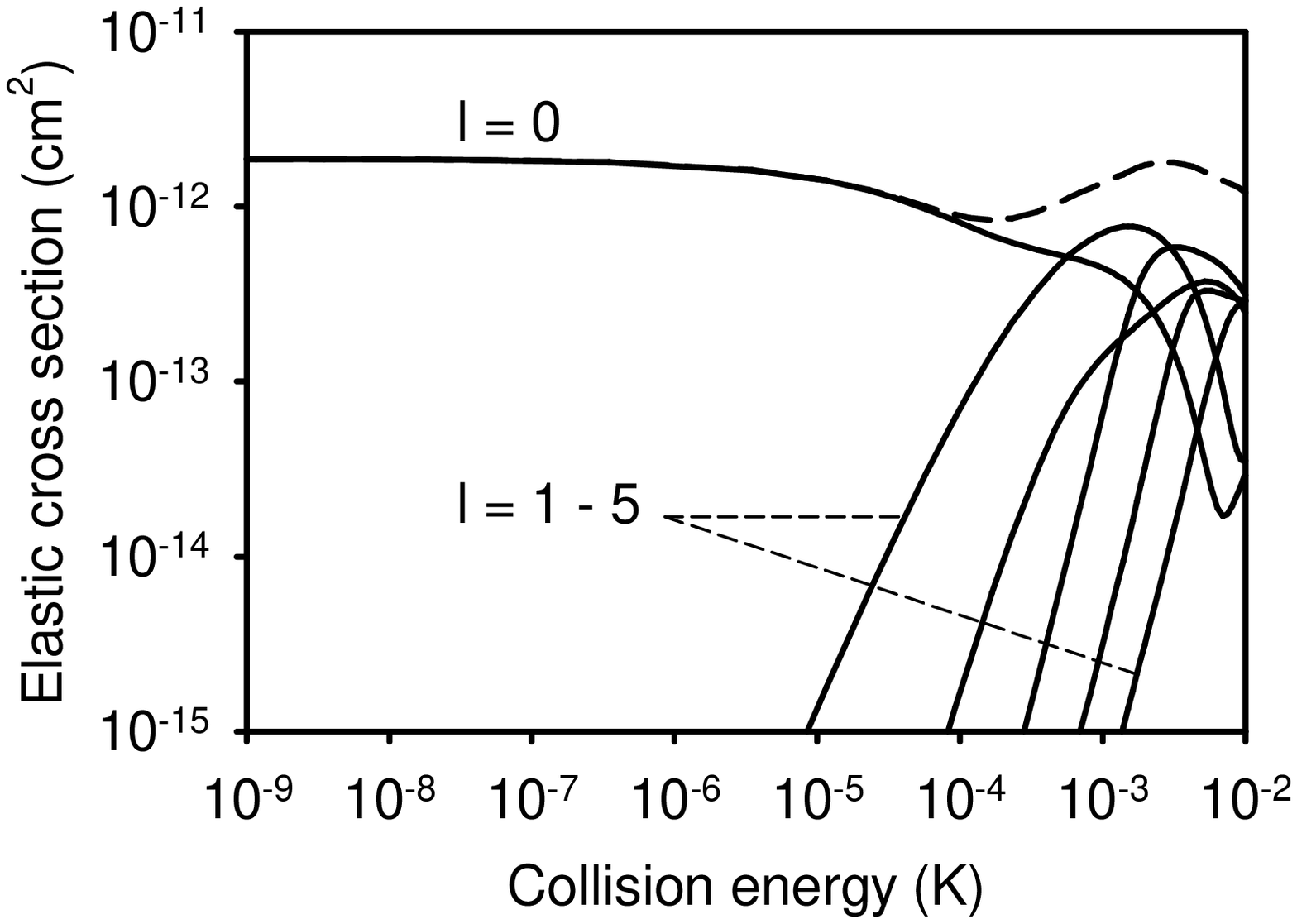}
\caption[Elastic cross section] {Elastic cross sections for
$^{39}$K + $^{39}$K$_2(v=1,j=0)$: partial as solid lines, total as
dashed line.} \label{ELCS}
\end{center}
\begin{center}
\includegraphics*[height=6cm,width=8cm]{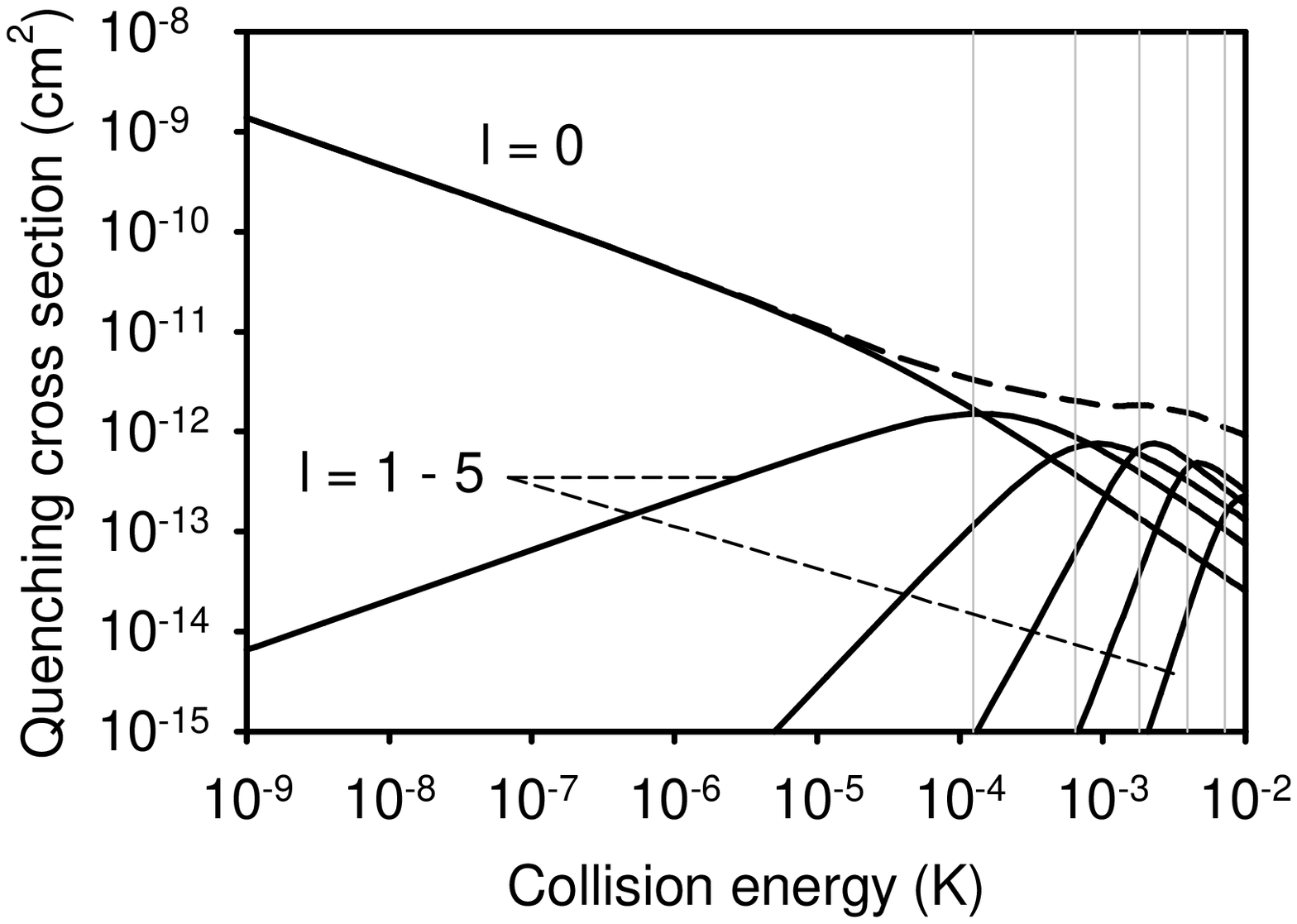}
\caption[Quenching cross section] {Quenching cross sections for
$^{39}$K + $^{39}$K$_2(v=1,j=0)$: partial as solid lines, total as
dashed line. Vertical lines correspond to the maxima of the
effective potential (see text for detail).} \label{QUCS}
\end{center}
\end{figure}

\section{Dynamical results and discussion}


The following discussion refers to collisions involving $^{39}$K
bosons unless otherwise stated.

\subsection{Hyperspherical adiabatic energies}

The hyperspherical adiabatic energies for the lowest $250$
hyperspherical states with $\Omega=0$ are shown as a function of
the hyperradius $\rho$ in Fig.~\ref{SPAGK3} (not shown here because the eps file is too big). 
If we consider
$^{39}$K$_2$ molecules in their $(v=1,j=0)$ rovibrational state,
we have 15 hyperspherical states asymptotically open, and 235
asymptotically closed. There are two minima: the deepest
corresponds to the equilateral geometry, while the secondary
minimum at around 20 a$_0$ corresponds to the linear geometry. At
large $\rho$, the hyperspherical adiabatic energies tend to the
energies of the diatomic molecule K$_2$. The zero of energy is
taken as the dissociation limit to three atoms.

\subsection{Cross sections and rate coefficients}

\begin{figure}[h]
\begin{center}
\includegraphics*[height=6cm,width=8cm]{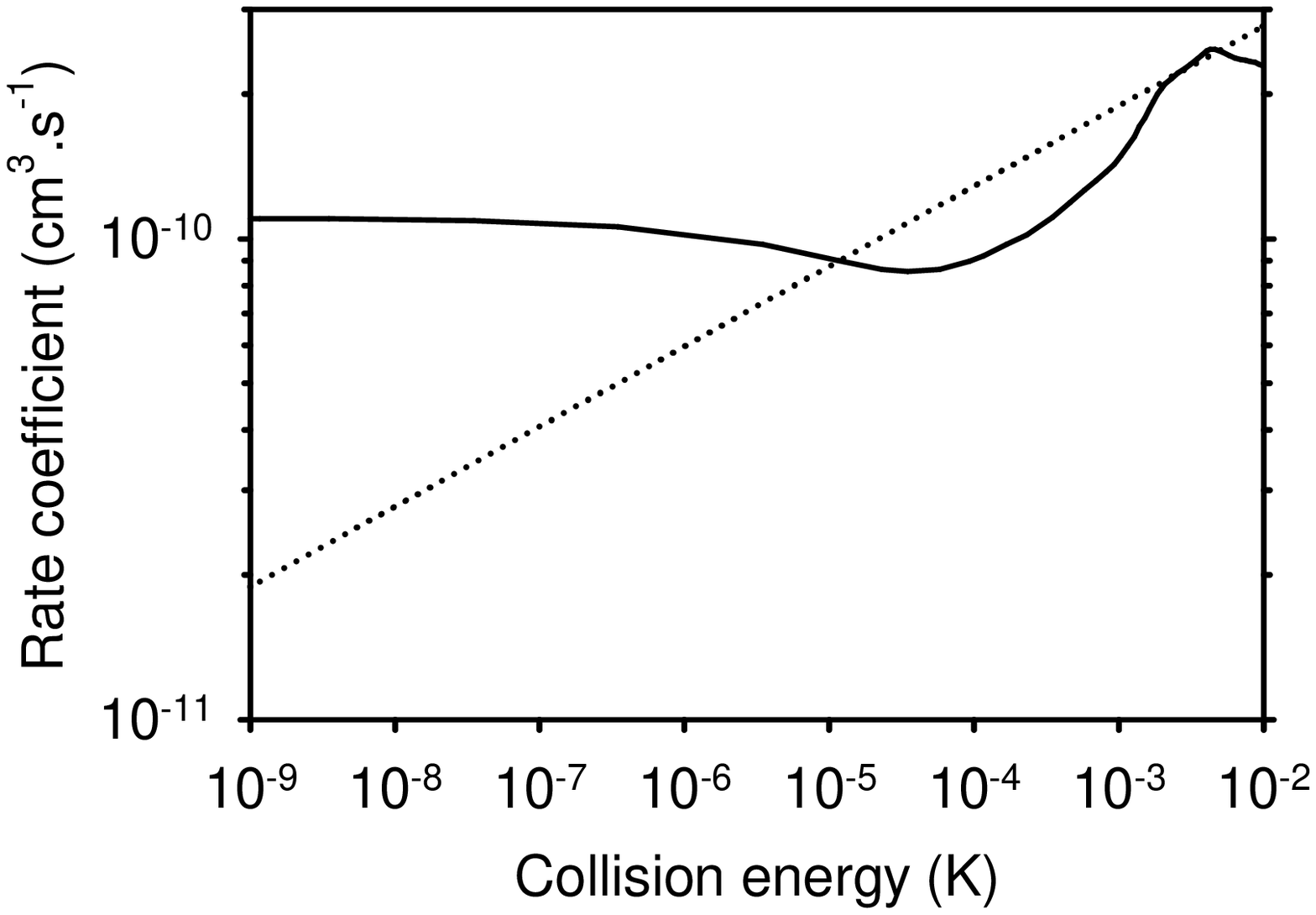}
\caption {Total quenching rate coefficients for $^{39}$K +
$^{39}$K$_2(v=1,j=0)$. The result from the Langevin model is shown
as a dotted line.} \label{RACO}
\end{center}
\end{figure}

The fully converged elastic and quenching integral cross sections
and their partial wave contributions for $l=0$ to 5 are shown in
Figs.\ \ref{ELCS} and \ref{QUCS} for collision energies between 1~nK and 10~mK.  
At $E_{\rm coll}=10^{-9}$ K, the elastic cross
section is about $1.9 \times 10^{-12}$ cm$^2$ whereas the
quenching cross section is about $1.4 \times 10^{-9}$ cm$^2$. The
slope of each partial cross section is given by the Wigner laws
(Eq.\ (\ref{wcross})) at ultralow energy 
except for the case of elastic scattering for $l>1$, where the 
dispersion-modified threshold law $\sigma_E^l  \sim E_{\rm coll}^3$ applies.
The upper limit of the
Wigner regime is 0.1 $\mu$K. The minimum in the $l=0$ partial
elastic cross section near 10$^{-2}$ K arises simply from a
near-zero phase-shift in the formula for the elastic cross
section.

The quenching cross section for a partial wave with $l>0$ shows a
maximum at a collision energy which is given approximately by the
maximum of the effective potential including the centrifugal and
dispersion terms,
\begin{eqnarray}
V^l(R)=\frac{l(l+1)}{2 \mu R^2} - \frac{C_{\rm K-K_2}}{R^6}.
\end{eqnarray}
The height of the barrier is
\begin{eqnarray}
V_{\rm max}^l=\frac{[l(l+1)]^{3/2}}{3 \mu^{3/2} (6 C_{\rm
K-K_2})^{1/2}}
\end{eqnarray}
at a distance
\begin{eqnarray}
R_{\rm max}^l=\left[\frac{6 \mu C_{\rm
K-K_2}}{l(l+1)}\right]^{1/4}.
\end{eqnarray}
The resulting barrier heights are included in Fig.~\ref{QUCS}. The
first vertical line corresponds to the $l=1$ partial wave and so
on up to the $l=5$ partial wave. It may be seen that each partial
wave has a maximum at an energy slightly higher than the
corresponding $V_{\rm max}^l$. At collision energies below the
centrifugal barrier, the quenching partial cross sections for each
$l$ follow Wigner laws given by Eq.\ (\ref{wcross}).  Above the
centrifugal barrier, the quenching probabilities come close to
their maximum value of 1 and the cross sections vary as $E^{-1}$
because of the $k^{-2}$ factor in the expression for the cross
section.

The total quenching rate coefficient is larger than the elastic
rate over a wide range of collision energies, up to 1 mK, and is
three orders of magnitude larger than the elastic rate at 1~nK
(Fig.\ \ref{BOSFERM}). The large quenching rates are consistent
with recent experiments that create molecules in atomic
Bose-Einstein condensates by Feshbach resonance tuning
\cite{Wiem02,Grimm03c,Remp04,Kett04}; for bosonic atoms, such
experiments were unable to produce a long-lived molecular
condensate because of the efficiency of quenching collisions.

At high collision energy, when several partial waves are involved
($l=0-5$), the total quenching rate coefficient can be compared
with that given by the classical Langevin capture model
\cite{levine87},
\begin{eqnarray}
K_{Q}^{\rm capture}(E)= \frac{3\pi}{2^{1/6}} \left(\frac{C_{\rm
K-K_2}^{1/3}}{\mu^{1/2}}\right)E^{1/6}
\end{eqnarray}
This rate coefficient is shown as a function of collision energy
in Fig.\ \ref{RACO}. It may be seen that there is
semi-quantitative agreement between our quantum results and the
capture model for collision energies above 0.1 mK. Below that
energy, differences appear because there are fewer partial waves
and the dynamics must then be described by a full
quantum-mechanical treatment.

\subsection{Differential cross sections}

\begin{figure}[h]
\begin{center}
\includegraphics*[height=6cm,width=8cm]{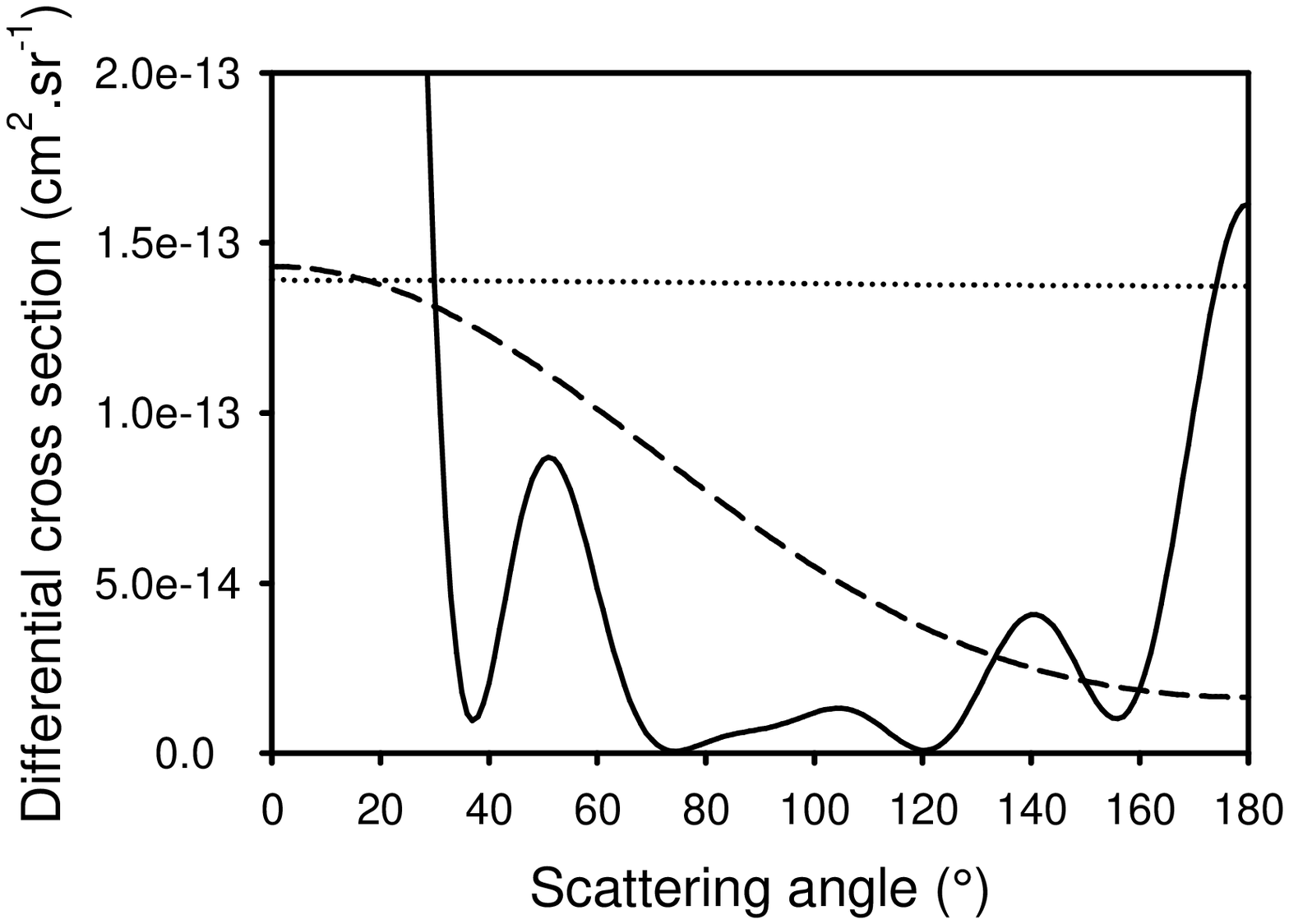}
\caption[DCSEL] {Differential cross section for elastic scattering
at 10$^{-6}$~K (dotted line), 10$^{-4}$~K (dashed line)  and
10$^{-2}$~K (solid line). The large forward scattering at
10$^{-2}$~K is not completely shown; at zero degree the DCS value
is $3 \times 10^{-12}$ cm$^2$sr$^{-1}$.} \label{DCSEL}
\end{center}
\end{figure}

\begin{figure}[h]
\begin{center}
\includegraphics*[height=6cm,width=8cm]{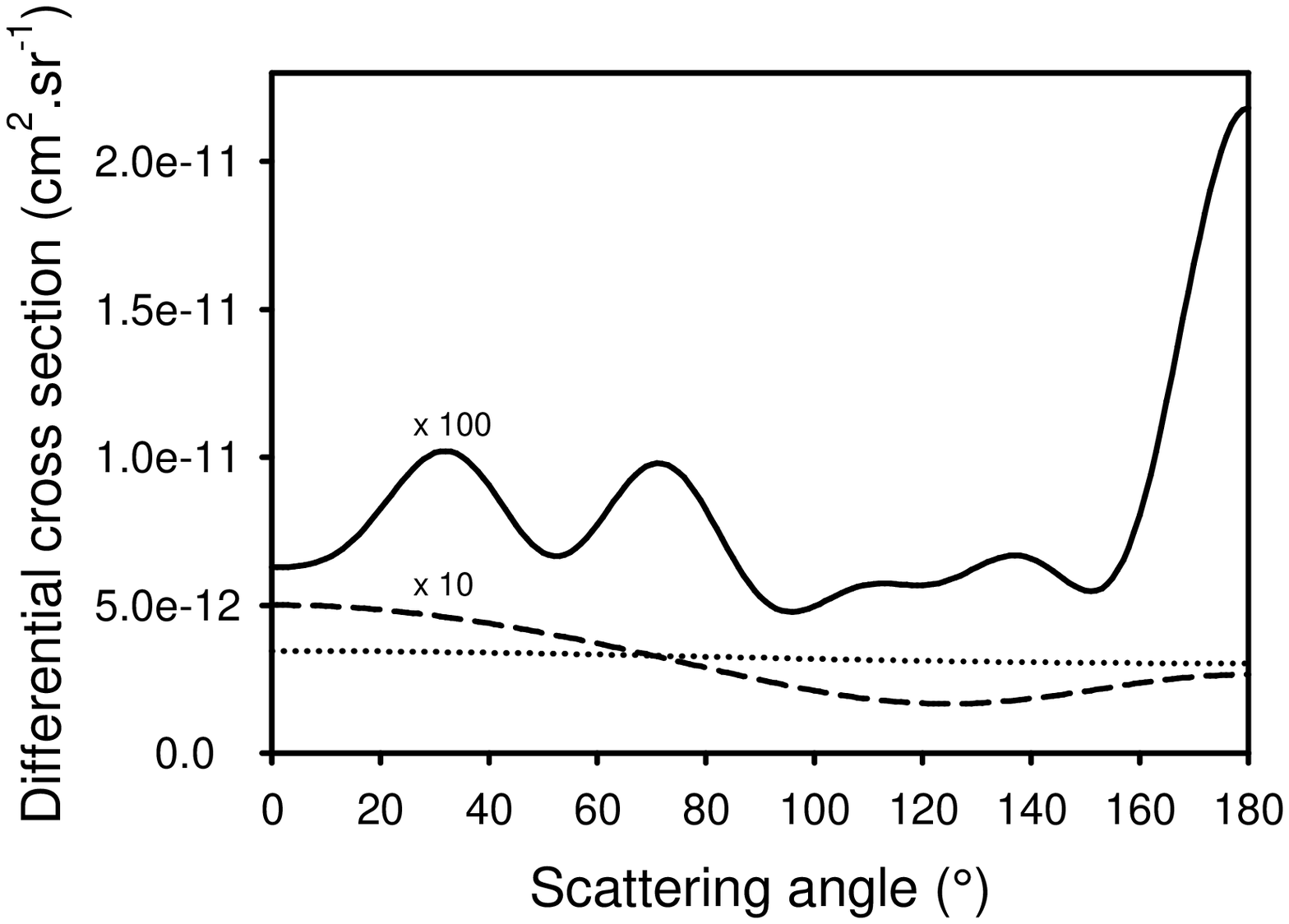}
\caption[DCSEL] {Differential cross section for vibrational
relaxation at 10$^{-6}$ K (dotted line), 10$^{-4}$ K (dashed line)
and 10$^{-2}$ K (solid line).} \label{DCSQU}
\end{center}
\end{figure}

Total differential cross sections (Eq. (\ref{dcs1})) for
elastic and quenching processes with $^{39}$K$_2$ molecules
initially in $(v=1,j=0)$ are shown in Figs.\ \ref{DCSEL} and
\ref{DCSQU} for collision energies 10$^{-6}$ K, 10$^{-4}$ K and
10$^{-2}$ K. At 10$^{-6}$ K, where only the $l=0$ and 1 partial
waves contribute, the elastic and quenching DCS are
quasi-isotropic. Below this energy (not shown), only the $l=0$
partial wave is involved and so the DCS are fully isotropic. At
higher energies, other partial waves contribute and the DCS depend
on the center-of-mass scattering angle $\theta_{\rm cm}$.

For the elastic process, Figure~\ref{DCSEL} shows that backward
scattering is smaller at 10$^{-4}$ K than at 10$^{-6}$ K, whereas
forward scattering is similar. At 10$^{-2}$ K, an enhancement of
forward scattering is found. Undulations at this energy are simply
due to a rainbow effect.

In contrast, the behavior of the quenching DCS is less systematic
(Figure~\ref{DCSQU}). The angular distribution is strongly peaked
in the backward direction at 10$^{-2}$ K whereas the forward
scattering is dominant at 10$^{-4}$ K. In addition, at 10$^{-2}$ K
the quenching DCS presents some oscillations in the sideways
scattering with two pronounced maxima at 30 and 70 degrees.

\subsection{Final rotational distributions}

\begin{figure}[h]
\begin{center}
\includegraphics*[height=6cm,width=8cm]{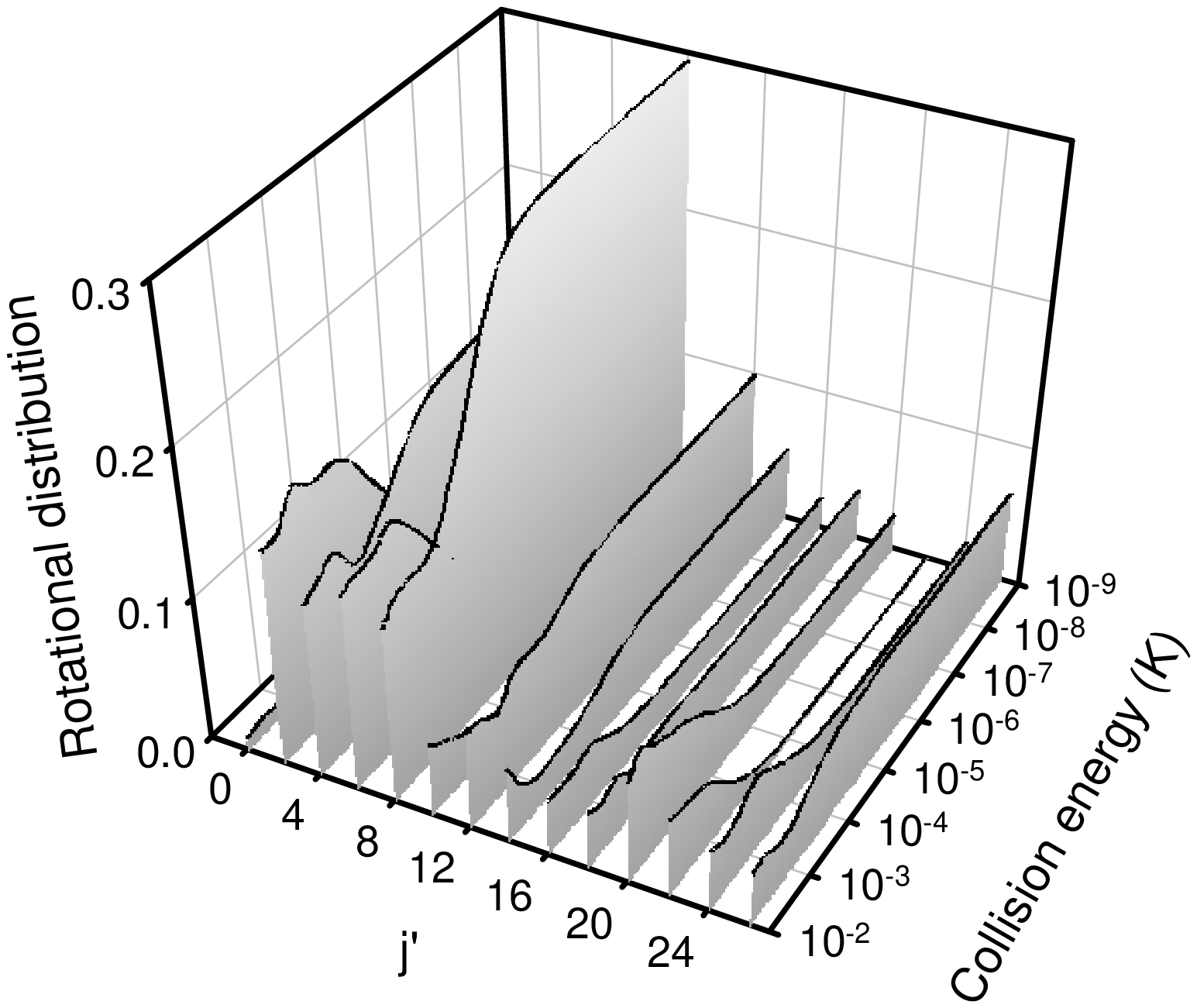}
\caption[Rotational distribution] {Rotational distributions as a
function of the collision energy. The label $j'$ is the final
rotational quantum number of $^{39}$K$_2(v'=0)$.} \label{ROTDIST}
\end{center}
\end{figure}

The final rotational distribution for $^{39}$K$_2$ molecules
initially in $(v=1,j=0)$ is shown in Fig.\ \ref{ROTDIST} as a
function of the collision energy. At each energy, we have divided
each rotationally-resolved cross section (corresponding to a given
final rotational quantum number $j'$ for $^{39}$K$_2(v'=0)$) by
the total quenching cross section. 
The sum of the distribution over $j'$ gives unity at fixed energy.
Because cross sections obey the Wigner laws (Eq.\ (\ref{wcross}), 
the rotational distribution becomes constant in the ultracold regime.
The final rotational state $j'=8$ is
the most populated at 1~nK whereas $j'=2,4,6,8$ are the most
populated at 10 mK. Figure \ref{ROTDIST} also shows that only the
lowest rotational states are significantly populated at all
collision energies.

\subsection{Bosons versus fermions}

\begin{figure}[h]
\begin{center}
\includegraphics*[height=6cm,width=8cm]{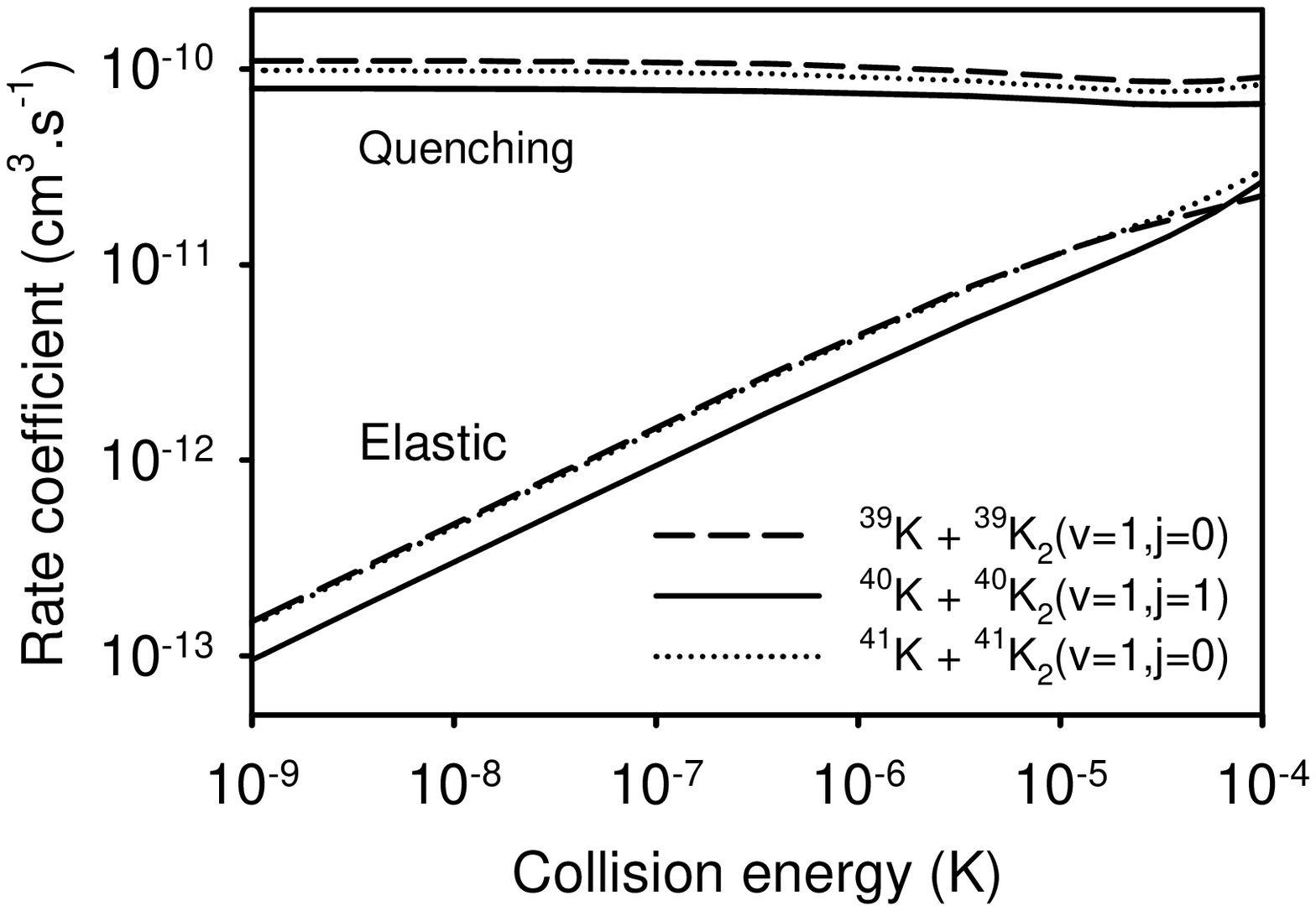}
\caption{Comparison of elastic and quenching rate coefficients for
bosons (dashed and dotted lines) and fermions (solid line).}
\label{BOSFERM}
\end{center}
\end{figure}

In Fig.~\ref{BOSFERM} we compare the total elastic and quenching rate
coefficients in the ultracold regime for the three systems:
$^{39}$K + $^{39}$K$_2$ or $^{41}$K + $^{41}$K$_2$, composed of
bosonic atoms and $^{40}$K + $^{40}$K$_2$, composed of fermionic
atoms.

In the ultracold regime, the total rate coefficients are very
similar for the two bosonic systems. However the rotationally
state-resolved cross sections (not shown here) are different and
the similarity of the total cross sections and rate coefficients is evidently
coincidental. The rotational distributions are sensitive to the
mass and also to the interaction potential as in Na + Na$_2$
\cite{Quem04}.

Figure \ref{BOSFERM} shows that the quenching processes are very efficient 
even for the fermionic system.
There is only a small
difference between fermions and bosons. Both elastic and quenching
rate coefficients are slightly smaller for the fermionic system than
for the bosonic systems. However the ratio of elastic to quenching
rate coefficients is nearly the same for bosons and fermions.

\section{Conclusion}
We have performed the first quantum dynamics calculations for K +
K$_2$ collisions in the energy range 1~nK - 10~mK, using a new
quartet potential energy surface for the potassium trimer. We have
found that the quenching rate coefficient is much larger than the
elastic rate coefficient at ultralow collision energies. This
applies for all three collisions studied here, $^{39}$K +
$^{39}$K$_2$, $^{41}$K + $^{41}$K$_2$, and $^{40}$K + $^{40}$K$_2$
where the initial K$_2$ molecules are in the $(v=1,j=0)$
rovibrational state for bosons or $(v=1,j=1)$ for fermions. Thus
K$_2$ molecules in these states are not good candidates to
accumulate large densities of ultracold molecules or to achieve
long-lived molecular BEC. For the bosonic systems ($^{39}$K and
$^{41}$K), the results are qualitatively similar to those we
obtained previously for $^{23}$Na \cite{Sol02,Quem04} and $^7$Li
\cite{Cvit04} atoms.

The quantum results for quenching rates agree semi-quantitatively
with the Langevin model at high collision energies, but below 0.1 mK
the dynamics is described only by the quantum theory.

The study of the fermionic system with $^{40}$K atoms of
particular interest, because it has recently been possible to
create stable molecular BECs of $^6$Li$_2$ and $^{40}$K$_2$
\cite{Jin03a,Grimm03a,Kett03a,Salomon04}. In these experiments,
molecules were formed in the highest vibrational state supported
by the potential well and it was found that inelastic collisions
were suppressed by Pauli blocking \cite{Gora03}. Our result shows
that for $^{40}$K$_2$ molecules in the initial vibrational states
$v=1$, there is no suppression of the quenching process.

Calculations with potassium atoms in higher vibrational states for
both bosonic and fermionic systems are in progress. Future studies
involving mixed collisions, i.e.\ with two different isotopes of
potassium, are also planned.

\section*{ACKNOWLEDGMENTS}
\label{secknow}

The dynamical calculations reported in this paper were performed
with computer time provided by the ``Institut du D\'{e}veloppement
des Ressources en Informatique Scientifique'' (IDRIS, Orsay) and
by the ``P\^{o}le de Calcul Intensif de l'Ouest'' (PCIO, Rennes).
PS and JMH are grateful to EPSRC for support under research grant
no.\ GR/R17522/01.

\end{document}